\documentstyle{mn}
\title{MERLIN/VLA imaging of the gravitational lens system B0218+357}
\author[A.~D.~Biggs et al.]{A.~D.~Biggs, \thanks{E-mail: adb@jb.man.ac.uk}
I.~W.~A. Browne, T.~W.~B.~Muxlow and P.~N.~Wilkinson\\
University of Manchester, Jodrell Bank Observatory, Macclesfield, 
Cheshire SK11 9DL}
\begin{document}
\maketitle
\begin{abstract}

Gravitational lenses offer the possibility of accurately determining the 
Hubble parameter ($H_0$) over cosmological distances, and B0218+357 is one of 
the most promising systems for an application of this technique. In particular 
this system has an accurately measured time delay ($10.5\pm0.4$~d; Biggs et al.
1999) and preliminary mass modelling has given a value for $H_0$ of 
69$^{+13}_{-19}$\,km\,s$^{-1}\,$Mpc$^{-1}$. The error on this estimate is now 
dominated by the uncertainty in the mass modelling. As this system contains an 
Einstein ring it should be possible to constrain the model better by imaging
the ring at high resolution. To achieve this we have combined data from MERLIN 
and the VLA at a frequency of 5~GHz. In particular MERLIN has been used in 
multi-frequency mode in order to improve substantially the aperture coverage 
of the combined data set. The resulting map is the best that has been made of 
the ring and contains many new and interesting features. Efforts are currently 
underway to exploit the new data for lensing constraints using the LensClean 
algorithm \cite{kochanek92}.

\end{abstract}


\section{Introduction}

The gravitational lens system B0218+357 \cite{patnaik93} has long been 
recognised as an excellent system for measuring the Hubble constant ($H_0$)
over cosmological distances using the method of Refsdal (1964). The system
has a simple morphology comprising two compact images (A and B) of a 
flat-spectrum radio core and a radio Einstein ring (Figs~\ref{mer1},
\ref{mer3} and~\ref{mervla}). There is also a kpc-scale radio jet that
extends south from the bottom of the Einstein ring (seen most clearly in 
Fig~\ref{vla}). Both lens and source redshifts are well determined, at 0.6847 
\cite{browne93} and 0.96 \cite{lawrence96} respectively. In addition, the time 
delay between the two compact components has been accurately measured 
($10.5\pm0.4$~d) from VLA monitoring \cite{biggs99}. Preliminary modelling of 
the mass distribution in the lensing galaxy using a Singular Isothermal 
Ellipsoidal (SIE) parameterization \cite{kormann94} gave a value for $H_0$ of 
69$^{+13}_{-19}$\,km\,s$^{-1}\,$Mpc$^{-1}$. The quoted error is a 95~per~cent 
confidence limit (statistical). Constraints on this model come from the VLBI 
substructure of the compact images (Patnaik, Porcas \& Browne 1995) as well as 
the flux density ratio measured from the VLA monitoring. A free parameter in 
the model is the position of the lensing galaxy centre which is poorly 
determined from {\em Hubble Space Telescope (HST)} optical and infra-red 
observations. Leh\'{a}r et al. (2000) have pointed out that the uncertainty 
in the position of the lensing galaxy implies an uncertainty in $H_0$ that is 
much greater than that given in Biggs et al. (1999).

A potential source of model constraints in B0218+357 is the radio brightness 
distribution in the Einstein ring which has been neglected until now. 
Einstein rings have proved to be a valuable source of constraints in other 
lens systems (e.g. MG~1131+0456 and PKS~1830-211) as they probe the mass 
distribution in the lensing galaxy at many points, thus providing many more 
constraints than are available from lenses that do not contain images of 
large-scale extended structure \cite{kochanek90}. Techniques that have been 
developed to optimise lens models using radio maps of Einstein rings include 
the Ring Cycle algorithm \cite{kochanek89} and LensClean \cite{kochanek92}. 

The Einstein ring in B0218+357 is believed to be an image of part of the 
extended emission of the kpc-scale radio jet. It is also the smallest known 
Einstein ring with a diameter of only 335~milliarcsec (mas), equal to the 
separation between the two compact components. Because of its small size, 
producing reliable maps of the brightness distribution in the ring that have 
high resolution and high sensitivity is difficult. The previous best 
map was made at a frequency of 5~GHz with the MERLIN array \cite{patnaik93}, 
but due to the incomplete aperture coverage of the observations no attempt 
was made to interpret details of the Einstein ring in a lensing context. This 
paper presents a new map of B0218+357 made from data obtained with both MERLIN 
(in multi-frequency mode) and the VLA at frequencies around 5~GHz. The 
excellent aperture coverage when these data are combined has allowed us to
produce the best map of this system at this frequency to date. Work is 
currently underway to exploit the new map for model constraints (Wucknitz et 
al., in preparation).

\section{Observations and Data Reduction} 

\subsection{MERLIN data}
\label{merlin}
Observations were taken in June 1995 with the MERLIN array in multi-frequency 
synthesis (MFS) mode (see Conway, Cornwell \& Wilkinson \shortcite{conway90} 
or Sault \& Conway \shortcite{sault99} for a full discussion of this 
technique). The advantage of this technique is that for a given baseline 
separation, changing the observing frequency changes 
the spatial frequency of the measured visibility and hence the position of the 
visibility in the $(u,v)$ plane. This allows the $(u,v)$ plane to be filled in 
more fully than would be possible for observations taken at a single frequency 
with the same array and covering the same period of time. This in turn allows 
the sky brightness distribution to be reconstructed more accurately. A
complication arises from the spectral index of the emission which, if 
uncorrected for, will produce imaging artefacts and significantly degrade the
dynamic range (the ratio of peak brightness to off-source error) of a map. 

For these observations three frequencies around 5~GHz 
separated by 320~MHz were used. These were 4546, 4866 and 5186~MHz, a 
frequency range of $\pm$7~per~cent. The bandwidth of the individual 
frequencies was 16~MHz. B0218+357 was observed alternately with the nearby 
JVAS calibrator source B0233+359 \cite{patnaik92} for phase-referencing 
purposes, a complete cycle of these two sources through all frequencies taking 
about half an hour before being repeated. Observations were taken irregularly 
over the period 17--20 June totalling approximately 36 hours of data. Full 
Stokes parameters were observed so that polarization maps could be made.

Initial data reduction followed standard procedures for MERLIN data. After
the flagging of corrupted data, the flux density scale was set relative to 
3C~286. Subsequent calibration was accomplished using the NRAO Astronomical
Image Processing Software ({\sc aips}). The phase-calibrator was found to be 
substantially resolved and was therefore iteratively mapped and self-calibrated
several times to produce an accurate model of the source structure. This was 
then used to derive antenna-based complex gain solutions which were then 
applied to the B0218+357 data. B0233+359 was also used to derive instrumental 
polarization corrections and 3C~286 was used to calibrate the polarization 
position angle.

\begin{figure}
\begin{center}
\setlength{\unitlength}{1cm}
\begin{picture}(8,8)
\put(-0.6,-2.1){\includegraphics{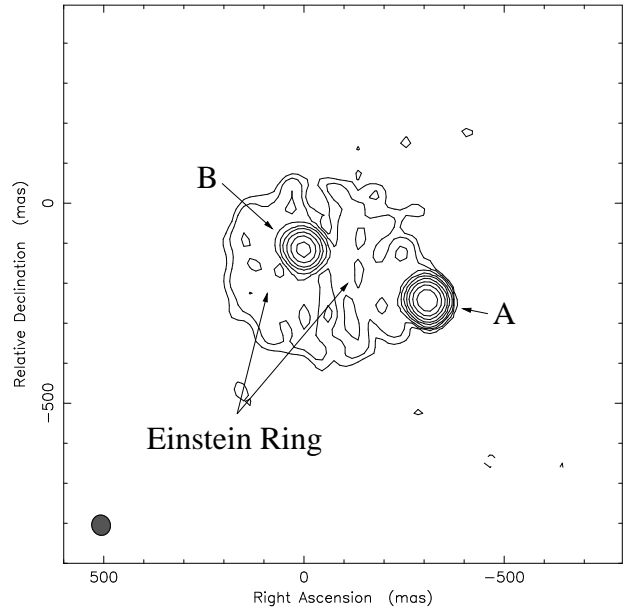}}
\end{picture}
\caption{MERLIN single-frequency (4546~MHz) map made using the {\sc difmap} 
program. The data are uniformly weighted and contours are plotted at $-0.2$,
0.2, 0.4, 0.8, 1.6, 3.2, 6.4, 12.8, 25.6 and 51.2~per~cent of the peak 
brightness in the map ($759\,\mathrm{mJy\,beam}^{-1}$). The restoring beam of 
$51\times47$~mas is shown in the bottom-left corner.}
\label{mer1}
\end{center}
\end{figure}

The data from each frequency were next individually mapped and 
self-calibrated using the {\sc difmap} package \cite{shepherd97}, solving for 
both phase and amplitude antenna corrections. A map made from the 4546~MHz 
data is shown in Fig.~\ref{mer1}. Before the three frequencies could be 
combined into a single data set, it was necessary to correct the data for the 
steep spectral index of the Einstein ring ($\alpha = -0.6$, where 
$S_{\nu} \propto \nu^{\alpha}$). Simply scaling each frequency's data was not 
possible due to the compact cores having a significantly flatter spectral 
index than the ring ($\alpha = -0.2$). This problem was solved by subtracting 
the compact cores from the $(u,v)$ data. This was achieved by (in {\sc aips}) 
making a map of the calibrated data output from {\sc difmap} and subtracting 
the Fourier transform of the CLEAN components corresponding to the positions 
of components A and B. The data were then remapped and the process repeated 
until all the core emission had been removed. The total flux density that 
remained in each subtracted map was then used to scale the $(u,v)$ data sets 
to a consistent amplitude scale. This correction was as expected, the higher
frequencies having to be scaled upwards in amplitude relative to the lower.
Finally, the CLEAN components removed from the central frequency were added 
back into the combined data set resulting in a `complete' B0218+357 data set. 

\begin{figure}
\begin{center}
\setlength{\unitlength}{1cm}
\begin{picture}(8,8)
\put(-1.4,8.2){\includegraphics{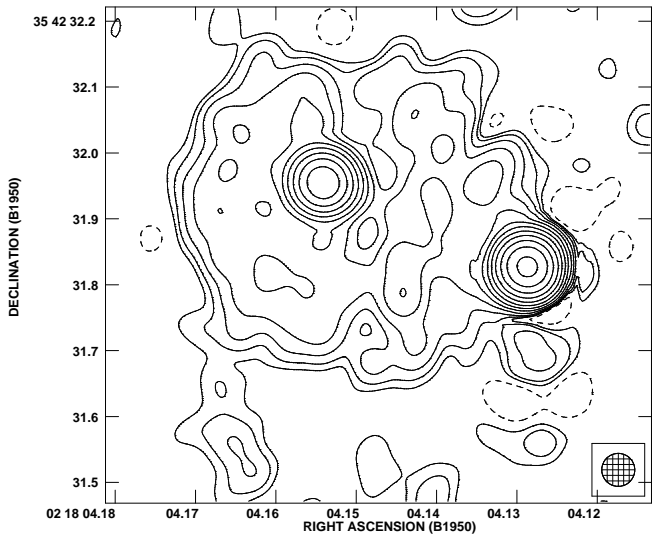}}
\end{picture}
\caption{MERLIN 5-GHz MFS map. Contours in this and subsequent maps are 
plotted at $-3$, 3, 6, 12, 24, 48, 96, 192, 384, 768, 1536 and 3072 times the 
off-source rms noise in the map ($102\mu\mathrm{Jy\,beam}^{-1}$). The circular 
restoring beam of $50\times50$~mas is shown in the bottom-right corner.}
\label{mer3}
\end{center}
\end{figure}

These data were then amplitude and phase self-calibrated (using the CLEAN 
components from an earlier 4546~MHz map) and mapped. In order to remove
spurious data, the brightest 1000 CLEAN components were vector-subtracted 
from the $(u,v)$ data and points with a residual amplitude of $>$200~mJy 
were removed from the data. This cut-off was chosen using a ``by-eye'' 
estimate of the scatter in the data, but as these only constituted
$<$0.7~per~cent of the data the effect on the resulting map is minimal.
The subtracted CLEAN components were then added back into the clipped data 
set, a map of which is shown in Fig.~\ref{mer3}. In this case the dirty map 
was made using {\sc mx} and CLEANed using the task {\sc sdcln}. This task 
utilises the Steer-Dewdney-Ito (SDI) method of choosing CLEAN components 
\cite{steer84} and is useful when a map contains extended structure of 
similar brightness. This map represents a considerable improvement on the 
single-frequency maps (such as that shown in Fig.~\ref{mer1}), the dynamic 
range having increased from 1500:1 to 8000:1. The map includes some low-level 
brightness from the jet to the south of the Einstein ring. There are also 
obvious miscalibration artefacts (alternate positive and negative sidelobes) 
around component A.

\subsection{VLA data}

\begin{figure}
\begin{center}
\setlength{\unitlength}{1cm}
\begin{picture}(8,8)
\put(-1.4,8.2){\includegraphics{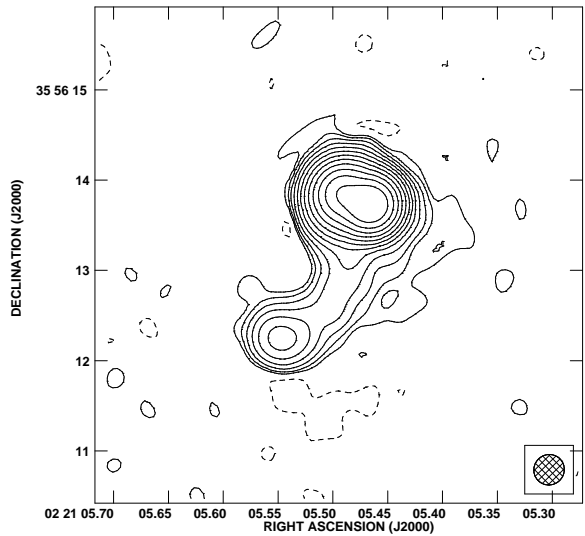}}
\end{picture}
\caption{VLA 5-GHz map. The data are uniformly weighted and the bottom 
contour is equal to three times the off-source rms noise 
($60\mu\mathrm{Jy\,beam}^{-1}$). The restoring beam of $341\times334$~mas is 
shown in the bottom-right corner.}
\label{vla}
\end{center}
\end{figure}

The VLA data were observed on the 8th August 1995 for a total of $\sim$100
minutes, less than two months after the MERLIN observations. Two separate
frequency bands were used and these were centred at 4885 and 4835~MHz, each 
with a bandwidth of 50~MHz. 3C~48 was used as the primary flux calibrator and 
3C~84 the point source phase-calibrator. Calibration followed standard 
procedures and proved to be straightforward with very little flagging of data 
necessary. B0218+357 was next iteratively mapped (using {\sc mx} with uniform 
weighting) and phase and amplitude self-calibrated to produce a map of the 
source which is shown in Fig.~\ref{vla}. This has an rms noise of 
60$\,\mu$Jy\,beam$^{-1}$, a figure that is substantially higher than 
the theoretical noise of $\sim$16$\,\mu$Jy\,beam$^{-1}$. This though is 
unsurprising given that the dynamic range in the map is $\sim$15\,000:1.
The map contains some likely spurious features (particularly emission to the 
west of component A and the jet), but these are only present at the 3$\sigma$ 
level, the lowest contour in the map. As the errors present in the VLA map are 
at such a low level of significance, they are unlikely to significantly 
degrade the quality of maps made from the combined MERLIN/VLA dataset.

\subsection{MERLIN/VLA combined data}

As the VLA data were observed in J2000 coordinates and MERLIN in B1950, the 
VLA data were transformed to B1950 coordinates. A map made from the
transformed $(u,v)$ data was essentially identical to that shown in 
Fig.~\ref{vla}. As a check on the accuracy of the coordinate transformation,
the task {\sc maxfit} was used to find the peak brightness and its position in 
both the MERLIN and the transformed VLA data sets, the results of which are 
shown in Table~\ref{maxfit}. 

\begin{table*}
\begin{center}
\caption{\label{maxfit} Positions and brightnesses of the brightest points
(component A) in the MERLIN (three-frequency) and VLA maps.}
\vspace{0.5cm}
\begin{tabular}{cccc}
Array   & Peak brightness (mJy\,beam$^{-1}$) & RA (hms) & Dec. 
($^{\circ}$\,$^{\prime}$\,$^{\prime\prime}$) \\ \hline\hline
MERLIN  & 933.2 & 02 18 04.12603 & +35 42 31.8348 \\
VLA     & 930.3 & 02 18 04.12998 & +35 42 31.8335 \\
\end{tabular}
\end{center}
\end{table*}

The VLA data were observed as close as possible in time to the MERLIN MFS data 
so as to keep to a minimum any complications in the combination of the two 
arrays arising from source variability. This has been successful as the peak 
brightness in each map is very similar. The positions of the peaks are a 
little less consistent. Whilst in declination the peaks are coincident to 
within 1~mas, the right ascension coordinates differ by 60~mas. Therefore 
the two maps are offset by about one MERLIN beam. This is not completely
surprising as no attempt was made to perform exact astrometry with the VLA
observations. The poor resolution of the VLA map may also be a contributing 
factor. However, as the offset in position is not unduly large (and removable 
with self-calibration) and the flux scales broadly consistent the two data 
sets were simply combined with no scaling or removal of model components (A 
and B). Prior to this the weights of each visibility of each array were made 
approximately the same so that each data set contributed roughly equally to 
the resultant image.

\begin{figure*}
\begin{center}
\setlength{\unitlength}{1cm}
\begin{picture}(10,18)
\put(-4,-3){\includegraphics{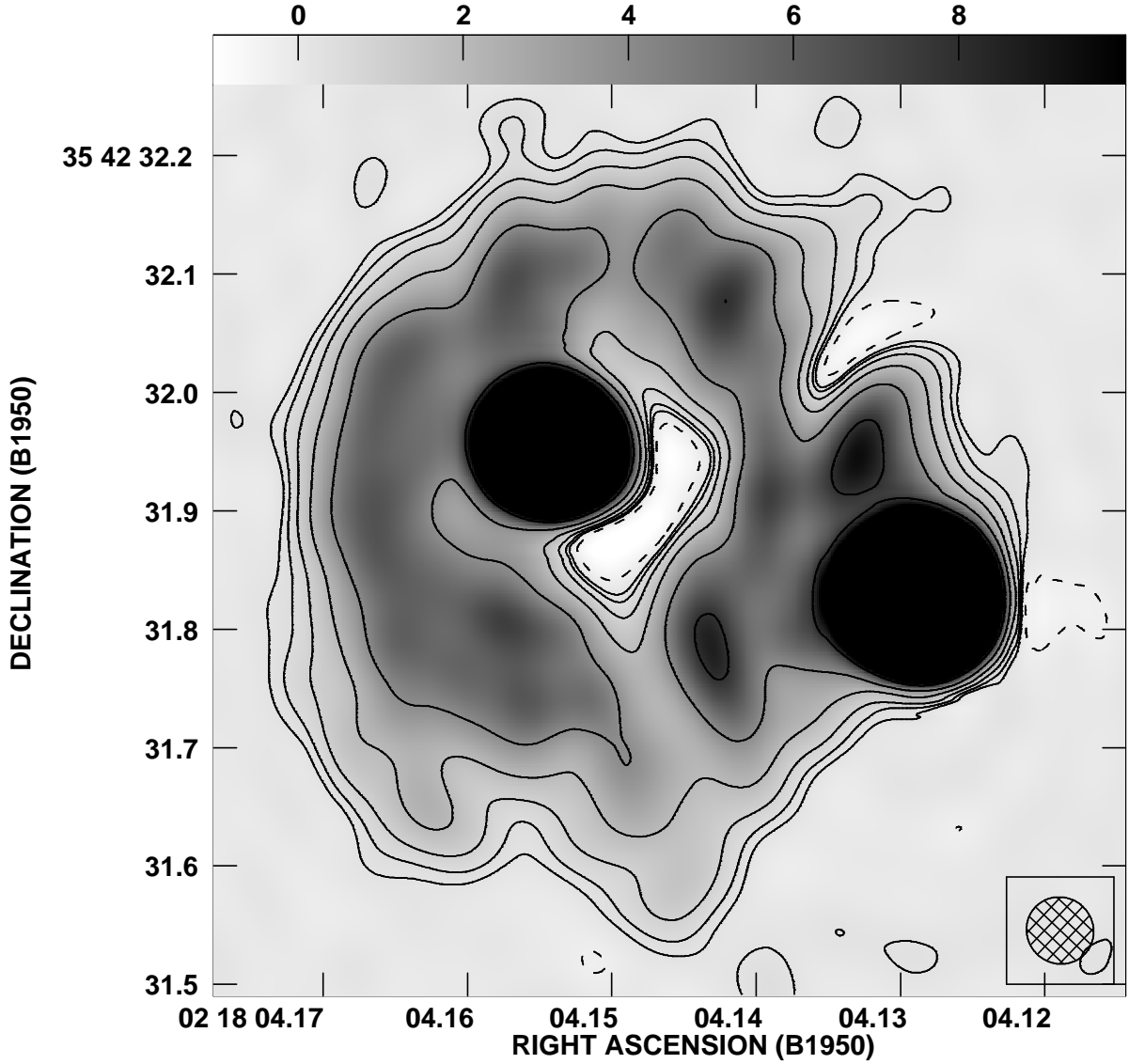}}
\end{picture}
\caption{MERLIN/VLA 5-GHz map. The data are uniformly 
weighted with a {\sc robust} parameter of $-1$. The bottom contour is equal to 
three times the off-source rms noise ($82\mu\mathrm{Jy\,beam}^{-1}$). 
Greyscales represent brightnesses between $-1$ and 10\,mJy\,beam$^{-1}$. The 
restoring beam of $58\times56$~mas is shown in the bottom-right corner.}
\label{mervla}
\end{center}
\end{figure*}

All maps of the combined data set were made using the {\sc aips} imaging task
{\sc imagr}. A value for the {\sc robust} parameter (which allows a compromise
to be made between the traditional natural and uniform weighting schemes) of
$-1$ was used in all maps which resulted in a beamsize of $57\times55$~mas.
The initial map of the combined data was very poor, but with several iterations
of phase self-calibration subsequent maps were of much higher quality. In order
to make the best possible map the data were also amplitude self-calibrated and
corrected for baseline errors. This latter step was particularly successful in
removing the sidelobe structure around component A seen in Fig.~\ref{mer3}.
The final image is shown in Fig.~\ref{mervla} and has an rms noise of 
$82\,\mu\mathrm{Jy\,beam}^{-1}$ and a dynamic range of 10\,000:1. The dynamic
range of a typical bright area of the ring is about 100:1.

\section{Discussion}

\subsection{Image fidelity}

The final image shown in Fig.~\ref{mervla} represents a marked improvement
on previous maps made of the Einstein ring in this lens system, combining the
sensitivity of the VLA with the resolution of MERLIN. As the dynamic range of
the map is greater and the aperture coverage so much better than in previous 
high-resolution maps, we can also expect there to have been a substantial 
improvement in the image fidelity (fractional on-source errors). Short of
performing complicated and time-consuming simulations of the entire mapping
process on a model source it is difficult to calculate the image fidelity. 
Instead, in the following paragraphs we will consider several ways in which 
the theoretical image fidelity could be degraded and show that the magnitudes 
of these are negligible.

The fundamental problem with MFS that must be overcome is that source 
brightness varies with frequency. We have compensated for this with the 
MERLIN data, as described in Section~\ref{merlin}, by removing the 
flat-spectrum cores and scaling the remaining emission. In doing so we have
assumed that the Einstein ring emission is described by a single value of 
$\alpha$. As gravitational lensing is an achromatic process this is in general 
a reasonable approach, providing that the lensed source has a uniform spectral 
index. If this is not the case then errors will result from the spectral-index 
residuals. We believe that the assumption of uniform $\alpha$ holds fairly 
well in B0218+357 due to the fact that the area of jet imaged into the ring 
is small (of order 10~mas) and because spectral-index gradients along radio 
jets are shallow \cite{bridle84}. Furthermore, Conway et al. (1990) have 
shown that for a knotty jet (of which the ring in B0218+357 could be 
considered an example with extreme curvature) observed with the MERLIN array 
with a bandspread of $<$25~per~cent, the spectral errors can usually be 
ignored when the dynamic range in the map is $<$1000:1. Our observations 
easily fulfill this criterion as the dynamic range of the ring emission is
$\sim$100:1 and the bandspread $\pm$7~per~cent. The much brighter cores do 
not contribute to the spectral sidelobes as after their subtraction at all 
three frequencies, only those from the central frequency were subsequently 
returned to the combined data. As the average VLA and central MERLIN 
frequencies differ by less than 1~per~cent and the peak brightnesses in the 
three-frequency MERLIN map and VLA map were so similar, it is unlikely that 
major spectral errors could result from the addition of the VLA data.

Another effect that will reduce the image fidelity is source flux density
variability. This needs particular consideration with regards to B0218+357 
as the radio core of the background source is variable, as it had to be for
the time delay to be measured. Fortunately, the relatively low frequency of
these observations means that any source variability should be reduced 
compared with the rapid variations seen at higher frequencies. VLA monitoring
data at 5,\footnote{Although only the 8.4 and 15~GHz data featured in Biggs
et al. (1999), a small amount of time per epoch was spent observing at 5~GHz.} 
8.4 and 15~GHz \cite{biggs99} show that although highly variable
at the highest frequency, the variations become much reduced in magnitude 
(by a factor of about one half) and have a longer timescale at 5~GHz. This 
theoretical argument is supported by the very similar peak flux densities 
in the MERLIN and VLA maps of this work despite their being observed 
approximately a month and a half apart.

\subsection{The new image}
\label{interpretation}
Perhaps the most striking feature of Fig.~\ref{mervla} is the 
absence of emission in the middle of the ring lying very close
to, and to the west of, component B. Related to this is the reduction in
brightness that forms a valley running north-south through the ring and 
passing through this hole. A similarly-aligned feature was seen at low 
significance in the independently-mapped 5-GHz MERLIN image of Patnaik et al. 
(1993), lending confidence in the reliability of this feature. The presence of 
this valley results in the ring taking on more the appearance of two arcs. 
Further support for the reduction in brightness close to component B comes 
from the 15-GHz VLA map shown in Biggs et al. (1999). The radio jet can be 
seen emerging from the southern edge of the Einstein ring as the two 
protrusions at the bottom. These correspond well to similar features in the 
VLA 15-GHz radio map of Biggs et al. (1999).

Another interesting area of the map is the bright spot (the third brightest 
component in the map) that lies outside the Einstein ring, $\sim$150~mas to 
the north of component A. This has not been identified before and is
noticeable by its absence in the MERLIN 5~GHz map of Patnaik et al. (1993). 
Recently though, a feature at the same position and with similar brightness 
has been seen in a combined MERLIN/European VLBI Network (EVN) image at 
1.7~GHz (A.~R.~Patnaik, private communication). In addition to this, 
four-telescope VLBI data (Vermeulen et al., in preparation) shows HI 
absorption associated with the lensing galaxy against three components --- 
A, B and a third component located close and to the north-east of A. For the
purposes of this discussion we shall refer to this as component C. What this 
third component might be is currently undetermined, but given that its 
position is so close to component A it is unlikely that it is not a lensed 
image. Its counterpart (component D say) is most likely to be located close to 
component B, but if we assume that the relative magnification between C and 
D is similar to that between A and B, component D will be difficult to detect.
This is because it will be weakened by about a third compared to C and closer 
to component B than C is to A by a similar factor.

It is the lensed emission in the Einstein ring that interests us most as it is
this which will enable us to constrain the lens model. One way of doing this is
to identify lensed images of the same surface brightness as these are likely 
to be images of the same background source (as gravitational lensing preserves
surface brightness). This is the basis of the Ring Cycle algorithm 
\cite{kochanek89} that attempts to optimise the mass model so that, when 
back-projected into the source plane, the surface brightness dispersion within 
individual pixels is minimised. A natural consequence of this process is a 
reconstruction of the unlensed source. However, it is clear from 
Fig.~\ref{mervla} that the two arcs of the Einstein ring contain substructure 
with different surface brightnesses. This is almost certainly due to the fact 
that the finite beam of the combined observations is not sufficient to resolve 
fully the brightness variations in the Einstein ring. In this case the 
appropriate action is to use the LensClean \cite{kochanek92} algorithm.
This represents a considerable improvement on the Ring cycle and uses the
CLEAN algorithm familiar to radio astronomers to allow for resolution effects
like those present in the combined MERLIN/VLA map of B0218+357. As with the
Ring cycle, LensClean simultaneously finds the best-fit lens model and
produces a map of the unlensed source structure. This work is currently 
underway and will feature in a future paper (Wucknitz et al., in preparation).

\section{Conclusions}

In this paper we have presented a new image of the gravitational lens system 
B0218+357 at a mean frequency close to 5~GHz. This has combined data from both 
the VLA and MERLIN, the latter in multi-frequency mode observing at three 
separate frequencies. The resulting data set has excellent $(u,v)$ coverage and
sensitivity, as well as resolution high enough ($\sim50$~mas) to easily resolve
the Einstein ring. The new map is the best made of this system.

The motivation for making this map was to improve the lens model for this 
system and hence to improve the estimate of $H_0$. This currently stands at
69$^{+13}_{-19}$\,km\,s$^{-1}\,$Mpc$^{-1}$ \cite{biggs99}, but there are only
limited constraints on the mass model available from the VLBI substructure of 
the compact images and the centre of the lensing galaxy is poorly determined. 
The next stage will be to exploit the new image for lensing constraints. The
results of applying the Lens Clean algorithm will allow us to improve 
substantially the estimate of $H_0$ from the time delay.

\section*{acknowledgements}

ADB acknowledges the receipt of a PPARC studentship. We thank Olaf Wucknitz
for fruitful discussions and the anonymous referee for several suggestions 
that significantly improved this paper. This research was supported in part 
by the European Commission TMR Programme, Research Network Contract 
ERBFMRXCT96-0034 `CERES'. MERLIN is run by the University of Manchester as a 
National Facility on behalf of PPARC. The VLA is operated by the National 
Radio Astronomy Observatory which is a facility of the National Science 
Foundation operated under cooperative agreement by Associated Universities, 
Inc.

\end{document}